\documentstyle[twocolumn,eqsecnum,aps,epsf]{revtex}% Use this for style revtex

\newcommand{\beq}{\begin{equation}}
\newcommand{\eeq}{\end{equation}}
\newcommand{\ba}{\begin{array}}
\newcommand{\ea}{\end{array}}
\newcommand{\bea}{\begin{eqnarray}}
\newcommand{\eea}{\end{eqnarray}}

\newcommand{\ket}[1]{| #1 \rangle}
\newcommand{\bra}[1]{\langle #1 |}
\newcommand{\proj}[1]{\ket{#1}\! \bra{#1}}
\newcommand{\inner}[2]{ \langle #1 | #2 \rangle}

\newcommand{\dubbelZ}{{\sf Z}\kern-.48em{\sf Z}}

\begin{document}
\draft
\title{Single quantum querying of a database}

\author{Barbara M. Terhal$^{(1)}$ and John A. Smolin$^{(2)}$} 
\address{$^{(1)}$Instituut voor Theoretische Fysica, Universiteit van Amsterdam, 
Valckenierstraat 65, 1018 XE Amsterdam,
and  Centrum voor Wiskunde en Informatica,\\
Kruislaan 413, 1098 SJ Amsterdam, The Netherlands.\\
Email: {\tt terhal@phys.uva.nl} \\
$^{(2)}$IBM Research Division, T.J. Watson Research Center, Yorktown Heights, New York 10598, USA.\\
 Email: {\tt smolin@watson.ibm.com} \\
}
\date{\today}
\maketitle

\begin{abstract}
We present a class of fast quantum algorithms, based on 
Bernstein and Vazirani's parity problem,  that
retrieve the entire contents of a quantum database $Y$
in a single query.  The class includes binary 
search problems and coin-weighing problems.
Our methods far exceed the efficiency of classical algorithms which
are bounded by the classical information-theoretic bound.
We show the connection between classical algorithms based
on several compression codes and our quantum-mechanical method.

\end{abstract}
\vspace{.1 in}
\vspace{.2 in}
\narrowtext

\section{Introduction}

Quantum computers have been shown recently to be able to 
solve certain problems faster than any known algorithm running
 on a classical computer \cite{simon,shor,grover}.  These
problems include factoring, which can be performed in polynomial 
time on a quantum computer \cite{shor},
but is widely believed to be exponentially difficult 
on a classical computer, and database lookup, which is 
provably faster on a quantum computer \cite{grover}.  
Understanding the power of quantum algorithms and 
developing new algorithms is of major interest as 
the building of a quantum computer will require a huge
investment.

In this paper we present quantum algorithms for binary search
and coin-weighing problems in which the information in a quantum 
database is retrieved with a single query.  
These are applications of Bernstein and Vazirani parity problem \cite{bv1}
and provide a strong illustration of the power of 
quantum computation and point out the limitations of classical
information-theoretic bounds applied to quantum computers.

Information theory is a useful tool for analyzing
the efficiency of classical algorithms.  Problems involving 
information retrieval from a database are
particularly amenable to such analysis.  
Consider this database search problem: we have a database $Y$
that contains $n$ items, of which a single one is marked. This database 
is represented as a bit string $y$ of length $n$ with Hamming weight one 
($y$ has exactly one ``1'').
One would like to locate the marked item in as few queries to the 
database as possible.
The queries are bit strings $x$ of length $n$ such that the database
returns the answer
\beq
a(x,y)=x\cdot y \equiv \left(\sum_{i=1}^n x_i y_i\right)\ {\rm mod}\ 2
\label{answer}
\eeq
where $x_i$ and $y_i$ are the $i^{\rm th}$ bits of $x$ and
$y$.  
A simple version of this problem is the case in which the
allowed queries $x$ have Hamming weight $1$.
The information retrieved by a single query $x_j=\delta_{ij}$ 
is small--it adds or eliminates item $i$ from the set of 
possible marked items. It thus takes $n-1$ queries to locate 
the marked item in the worst case. A surprising 
result of Grover~\cite{grover} is that a quantum mechanical 
algorithm can be faster than this and find the marked item 
with high probability in $O(\sqrt{n})$ 
{\em quantum} queries, contrary to one's ``classical''
intuition.   Grover's algorithm does not, however,
violate the information theoretic lower bound on the minimal number of
queries $M$.

The information-theoretic lower bound \cite{aigner} on $M$ is given by
the amount of information in the database divided by the maximal amount
of information retrieved by a query which has $A$ possible answers, i.e.
\beq
M \geq\frac{H(Y)}{\log_2{A}},
\label{infobound}
\eeq
where $H(Y)=-\sum_y p_y \log_2{p_y}$, $p_y$ is the probability 
for $Y$ to contain $y$ and $\sum_y p_y=1$.  

A quantum algorithm employs a database which responds to superpositions 
of queries with superpositions of answers.   The quantum database
acts on two input registers:  register $X$ containing the query 
state $\ket{x}$ and register $B$, an output register
of dimension $A$ initially containing state $\ket{b}$. 
We define the operation of querying the database as 
\beq
R_y\colon\; \ket{x,b} \rightarrow 
\ket{x, [b + a(x,y)] \ {\rm mod}\ A} 
\label{requation}
\eeq
where $R_y$ is a classical reversible transformation which maps
basis states to basis states (that is, a 
permutation matrix) depending on the contents of the database, 
and $a(x,y)$ is the answer to query $x$, given database state $y$.
In a classical query only query basis states
$\ket{x}$ are used and the output register $B$ is 
initially set to $\ket{0}$.  However, a quantum database is not 
restricted to working only on basis states but can handle 
arbitrary superpositions of inputs \cite{nonuni}.
Because of this the information that is retrieved by a single quantum query is 
not bounded by $\log_2 A$.
The relevant quantity in the quantum setting is the accessible 
information in the registers $X$ and $B$ (together called $XB$) and the 
internal state of the quantum computer $\Phi$ about the database $Y$. 
Together these are always in a one of a set of pure states
$\{\ket{\psi_y},p_y\}_{y \in Y}$ and the accessible information 
on $y$ is bounded by the Kholevo bound \cite{hol}
\beq
I_{\rm acc}(\Phi XB) \leq S(\Phi XB),
\eeq
where $S(\Phi XB)=-{\rm Tr }\rho_{\Phi XB} \log_2 \rho_{\Phi XB}$ is the Von 
Neumann entropy of $\Phi XB$ and $\rho_{\Phi XB}=\sum_y p_y \proj{\psi_y}$.
In the case of a classical query, the Von Neumann entropy 
$S(\Phi XB)$ is strictly less than $\log_2 {\rm dim}(B)=\log_2 A$ 
which gives rise to the classical bound (\ref{infobound}). 
%In the case of 
%a quantum query, $S(XB)$ is strictly bounded by $\log {\rm dim}
%(XB)$. 
The quantum algorithms that will be presented in this paper ``violate'' 
the classical information-theoretic bound 
by extracting extra information in the phases of the query register $X$.
It is notable that Grover's Hamming weight one problem \cite{grover}
has been proven  optimal \cite{strength,tight}; no quantum
algorithm for this problem can violate the classical information-theoretic bound (\ref{infobound}).

The quantum algorithms presented here are a major improvement on the classical 
algorithms
in terms of computation time if the computation performed by the database is costly.
All our algorithms make use of an interaction with the database of the form
$a(x,y)=x \cdot y$.  A direct implementation of such a database takes
$O(\log n)$ time using $n$ Toffoli and $n$ XOR gates in parallel.
The database can however be more general than this.  Any function 
$f(x)$ with the property that it can be written as $f(x)=x \cdot y$
can function as a database. The circuit that computes $f(x)$
for any input $x \in \{0,1\}^n$ runs in some time $T(n)$. We will 
compare the running time of the quantum algorithms including 
this circuit to the classical running time.

\section{The Parity Problem and Coin Weighing}

Bernstein and Vazirani \cite{bv1,bv2} have given the first problem in
which a single quantum query to the database is sufficient and a
strong violation of the classical information theoretic bound comes
about.  In their parity problem they consider a database $Y$ which
contains an arbitrary $n$-bit string $y$. The answer to queries
represented by $n$-bit strings $x$ to the database is the parity of
the bits common to $x$ and $y$ given by $a(x,y)=x \cdot y$.  Note the
the problem is to determine $y$ in its entirety, {\em not} to merely
determine the parity of $y$.  Bernstein and Vazirani show that $y$ can
be determined in a single query to the database.  Here we apply this
quantum algorithm to the coin weighing problem.

Coin weighing problems are a group of problems in which a set of 
defective coins is to be identified in a total set of coins. 
Assume there are two types of coins, good and bad ones, 
and we can weigh arbitrary sets of coins with a spring-scale (
which gives the weight of the set of coins directly, as opposed to a balance
which compares two sets of coins).
All sets of coins are equiprobable. A set of 
$n$ coins is represented as a bit string $y$ of length $n$ 
where $y_i=1$ indicates that coin $i$ is defective. A 
weighing can be represented by a query string $x$, 
where $x_i$ specifies whether coin $i$ is included in the set 
to be weighed.  The result of a classical weighing is the Hamming
weight of the bitwise product of $x$ and $y$, $w_H(x \wedge y)$.
For this problem the information theoretic bound (\ref{infobound}) gives
\beq
M \geq \frac{n}{\log_2(n+1)}.
\label{clas1}
\eeq
This is close to what the best predetermined algorithm which perfectly identifies
the set of coins can achieve \cite{aigner}
\beq
\lim_{n \rightarrow \infty} M_{\rm pre}(n)=\frac{2n}{\log_2(n)}.
\label{clas}
\eeq

If one has a spring scale capable of performing weighings in superposition, 
then, one can use the Bernstein-Vazirani algorithm to identify the 
defective coins perfectly with a single weighing. 

Define $n'\equiv 2 \lceil \frac{n+1}{2}\rceil -1$. We construct a query state
\beq
\ket{\psi}=\frac{1}{\sqrt{2^{n}}}\sum_{x} \ket{x}\otimes \frac{1}{\sqrt{n'+1}}
\sum_{b=0}^{n'}(-1)^b \ket{b}.
\eeq 
The special preparation of register $B$ will make the result of a 
quantum query end up in the phases of $X$ while leaving
register $B$ itself unchanged (cf. \cite{strength}). After the query we have (using $(-1)^{w_h(x \wedge y)}=(-1)^{x \cdot y}$)
\beq
\ket{\psi_y}=\frac{1}{\sqrt{2^{n}}}\sum_{x} (-1)^{x \cdot y} \ket{x} 
\otimes \frac{1}{\sqrt{n'+1}} \sum_{b=0}^{n'} (-1)^b \ket{b}.
\label{ortho3}
\eeq
Thereafter we perform a Hadamard transform $H$ on the query register
\beq
H\colon\;\ket{x} \rightarrow \frac{1}{\sqrt{2^n}}\sum_{z} (-1)^{x\cdot z} \ket{z}.
\eeq
This results in the final state 
\beq
\ket{y} \otimes \frac{1}{\sqrt{n'+1}} \sum_{b=0}^{n'} (-1)^b \ket{b},
\eeq
thus retrieving $y$ within a single query.

Note that this coin-weighing algorithm uses only the parity of 
the Hamming weight of the answer whereas the full Hamming weight is 
available from the database and is fruitfully used in the classical algorithm. 

We can compare the total running time of this quantum algorithm with
that of the classical algorithm.  The preprocessing and postprocessing
of the register $X$ of the query state and the preprocessing of the
$B$ register all consist of the Hadamard transforms on individual bits
\beq 
R=\frac{1}{\sqrt{2}}\left(\ba{cc} 1 & 1 \\ 1 & -1 \ea\right).
\label{had}
\eeq
which can be done in parallel.  The total running time is then
simply $2+T(n)$.
In the classical algorithm the database circuit 
is used at least (see Eq. \ref{clas1}) $n/\log_2 (n+1)$ times resulting in
a total running time of at least $n T(n)/\log_2 (n+1)$.

\section{Compressive algorithms}

In this section we will consider modifications of this problem
in which the information in the database $H(Y)$ is less than $n$ bits. 
In these cases retrieving the data from the source can be viewed 
as a problem of data compression of a source $Y$. 
We will restrict ourselves here to the compression of the data from 
a single use of the source. In the classical case, each query to 
the database retrieves a single digit
of the word into which the bit string $y$ will be encoded.
The minimal set of predetermined classical queries will serve to construct 
the single quantum query algorithm.
We will use coding schemes that minimize the amount of pre/postprocessing
in the single query quantum algorithm.
A classically optimal encoding scheme (cf. \cite{cover}) has
\beq
H(Y) \leq \sum_i p_i l_i \leq H(Y)+1,
\eeq
where $l_i$ the length of the compacted $y_i$ and $p_i$ the probability that
the database contains bit string $y_i$. It is not guaranteed however that 
such an optimal encoding scheme can be implemented by the type of
database interaction that one is given to use, namely (\ref{requation}).

In the following section we present single query quantum algorithms of which
the construction is based on optimal classical encoding schemes, namely Huffman
coding. In the section thereafter we consider more general type of databases
and use a random coding scheme.  Each of these schemes will require 
precomputation of a set of queries based on the encoding schemes.  The
time for this computation will not be counted in the total running
time as the queries can be precomputed once and reused on subsequent problems.

\subsection{Binary search problem}

Binary search problems are defined as problems in which the database 
responds with two answers to the query. Here we look at such a search
problem in which the queries have Hamming weight $n/2$. The database
contains a bit string $y$ with Hamming weight 1. Let us first look at
the problem in which all these bit strings are equiprobable. We assume 
that $n$ is an integer power of two. For other $n$ one simply extends
the database size to the next higher power of two.

Classically 
it is well known that the marked item can be found in $\log_2 n$
queries, which achieves the classical information-theoretic 
bound (\ref{infobound}). The $k^{\rm th}$ query, $g_k$, is a string of
$2^{k-1}$ zeros alternating with a string of $2^{k-1}$ ones, 
where $k=1\ldots\log_2 n$, i.e.
\beq
\ba{c}
g_1=01010101.. \\
g_2=00110011..  \\
g_3=00001111.. \\
\mbox{ etc. } \\
\ea 
\eeq 

%(i.e. the $i^{\rm th}$ bit of $g_k$, $(g_k)_i$, is the
%$j^{\rm th}$ bit of $i-1$ expressed in binary). 

The result of query $g_k$ is 
\beq
z_k \equiv g_k \cdot y = a(g_k,y),
\label{zk}
\eeq
where $z_k$ is the $k^{\rm th}$ bit of the encoding $z$ of $y$. 
Each $y$ will have a different encoding $z$ and thus $z$ 
uniquely determines $y$.
The $g_k$s are the generators of the group $F$ of Walsh 
functions whose group multiplication rule is 
addition modulo 2. We can represent a Walsh function $f_s$ as
\beq
f_s=\sum_{i=1}^{\log_2 n}g_i s_i\ {\rm mod}\ 2,
\eeq
where $s$ is an arbitrary bit string of length $\log_2 n$.
The quantum-mechanical algorithm makes use of superpositions of all the Walsh functions. 
We construct the query state 
\beq
\ket{\psi}=\frac{1}{\sqrt{n}} \sum_{s} \ket{s,f_s} \otimes \frac{\ket{0}-\ket{1}}{\sqrt{2}} .
\eeq
After one query the state becomes
\beq
\ket{\psi_y}=\frac{1}{\sqrt{n}} 
  \sum_{s} (-1)^{a(f_s,y)} \ket{s,f_s} \otimes \frac{\ket{0}-\ket{1}}{\sqrt{2}}.
\eeq
It can be shown that
\beq
\inner{\psi_{y}}{\psi_{y'}}=\frac{1}{n}\sum_{s} (-1)^{a(f_s,y)+a(f_s,y')} =\delta_{yy'}. 
\label{ortho1}
\eeq
We can write 
\beq
a(f_s,y)=\sum_{k=1}^{{\rm log_2 } n} s_k a(g_k,y)\ {\rm mod}\ 2.
\eeq
Using (\ref{zk}) it follows that $a(f_s,y)=s \cdot z$,
and with this we can rewrite (\ref{ortho1}) as
\beq
\inner{\psi_{y}}{\psi_{y'}}=\frac{1}{n}\sum_s (-1)^{s \cdot z + s \cdot z'}=
\delta_{zz'}=\delta_{yy'}.  
\label{ortho2}
\eeq

As all states $\ket{\psi_{y}}$ and $\ket{\psi_{y'}}$ are orthogonal, they can 
be distinguished by a measurement and no further queries to the database
are required. 

\begin{figure}
\epsfxsize=8.0cm
\leavevmode
\epsfbox{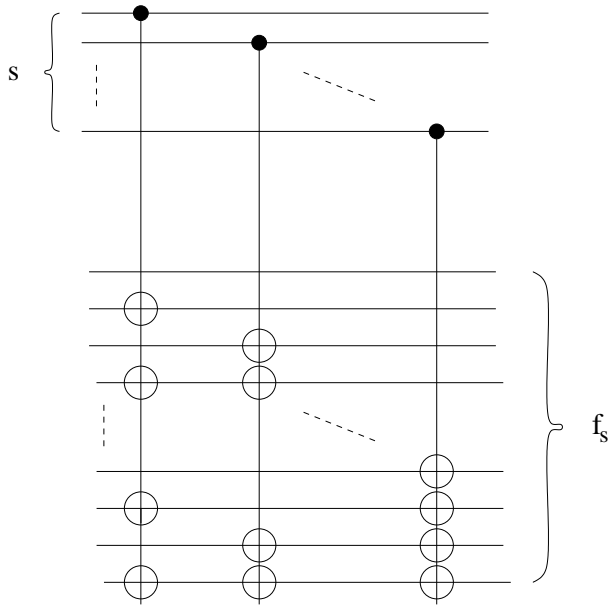}
\caption{``Walsh'' Circuit }
\label{fig2}
\end{figure}

What are the transformations that are required for pre- and
postprocessing of the query?  The preparation of the queries takes
$1+n/2 \log_2 n$ steps.  Register $s$ is prepared in superposition
using parallel one-bit Hadamard transforms and used as input to the
circuit shown in Figure \ref{fig2}.  The circuit in Figure \ref{fig2}
uses multi-bit XORs which we have counted as being in series.  The
same sequence in reverse is used as the postprocessing.  The total
time is thus $2+n \log_2 n + T(n)$.  In the classical case the queries
are also prepared using the circuit in Figure \ref{fig2}, but
the multi-bit XORs can be done in parallel, and the total
time is $ (\log_2 n)/2  + T(n) \log_2 n$.  
Note that in the Cirac-Zoller ion-trap model \cite{zoller} 
of quantum computation a multi-bit XOR gate can be done in parallel
by using the ``bus phonon'' modes.  Such quantum computers run our
algorithm in time $2+\log_2 n + T(n)$.

Note that we could have used all possible queries as in section II to
retrieve $y$ and subsequently compacted $y$ to $z$. This would have 
taken $2+T(n)$ for the algorithm plus  $n\log_2 n$ steps for the
compression, which is the same as this direct compression.

If we generalize this problem to databases that have unequal probabilities
assigned to different $y$'s, a scheme based on Huffman coding \cite{cover} is 
sometimes more efficient in terms of pre/post processing.
We assume that the probability distribution, or $H(Y)$ is known beforehand.

\begin{figure}[htbf]
\epsfxsize=8.0cm
\leavevmode
\epsfbox{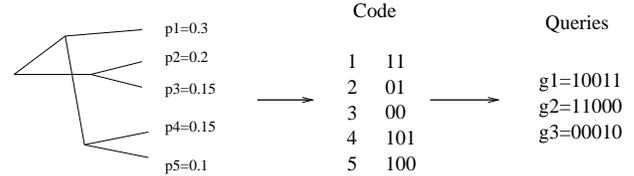}
\caption{Example of a Huffman code}
\label{fig3}
\end{figure}

Huffman coding is a fixed-to-variable-length encoding of a source, in
our case the database, which is optimal in the sense that it minimizes
the average codeword length $\sum_i p_i l_i \leq H(Y) +1$ with $H(Y)$
the entropy of the source.  The encoding prescribes a set of queries
that play the role of the Walsh generators in the equal probabilities
case. This is illustrated with an example in Figure \ref{fig3}. (In
fact, the Huffman construction results in Walsh queries in the
equal-probability-case.)

Classically, instead of querying with the Walsh generators, one can
use these Huffman queries, until the marked item has been found. The 
optimality of the Huffman code assures that the expected number of 
queries is minimized.
Choose the set of queries that will take the place of the Walsh generators
in the following way. Select a set of $m$ queries, the first $m$ queries
that are used in the classical case, such that the probability of not
finding the marked item after these $m$ queries is very small.
The value of $m$ will depend on the probability distribution.  If 
\beq
m \leq \lceil \log_2 n\rceil,
\eeq 
this Huffman scheme can be more efficient than a Walsh scheme. 

Note that this requirement is not necessarily satisfied for all 
probability distributions. For example for the distribution 
$p_1=1/10, p_i=\frac{9}{10 n}, i=1\ldots n$, the length of  
$n-1$ encoded words will be about $\lceil \log_2 n \rceil$. 
If we choose $\lceil \log_2 n\rceil-s$ queries,
the probability of error will be go to $9/10$ exponentially as
$2^{-s}$.

This set of $m$ queries will take the place of the Walsh generators in 
our quantum algorithm. The circuit which implements the Huffman queries
will be as in Fig. \ref{fig2} but with a different pattern of XORs
corresponding the Huffman queries.
All the database states that gave rise to 
distinct codewords after these $m$ classical queries will 
give rise to distinct $\ket{\psi_y}$ in our quantum algorithm.

If we query only once, the total running time will be $2+mn + T(n)$ if
we are willing to accept a small chance of error.
A classical algorithm that uses the same Huffman queries and has the
same probability of error takes $m/2+m T(n)$ time. Thus for some
probability distributions, $m$ can be significantly smaller than 
$\log_2 n$ and the algorithm is faster than a straightforward 
search with the Walsh queries.

\subsection{Random coding}
The binary search and the coin weighing problem are special cases
of a more general problem in which we have a database $Y$ that 
contains $k$ arbitrary base $A$ strings of length $n$.  Here we 
restrict ourselves to databases that contain $k$ equally probable strings.

The queries $x$ are all possible base $A$ strings of length $n$, the elements of 
$(\dubbelZ_A)^n$. The database returns the answer
\beq
a(x,y)=\sum_i^n x_i y_i\;  {\rm  mod}\; A \equiv x \cdot y\ .
\label{defans}
\eeq 
The information $H(Y)$ is equal to $\log_A k$. A classical
predetermined algorithm to determine $y$ with high probability 
makes use of $m=\log_A k+l$ random strings, where $l$ is a small integer. 
Pick $m$ linearly 
independent random base $A$ strings of length $n$; these are the
queries $g_i$, $i=1 \ldots m$. Similarly to (\ref{zk}) we define 
the encoding as
\beq
z_k=g_k \cdot y
\label{hash}
\eeq
where $z_k$ is the $k^{\rm th}$ digit of $z$. The $g_i$s are used 
to compress the string $y$ of length $n$ to the codewords 
$z$ of length $m$. What is the probability that the 
codeword $z$ determines $y$ uniquely? The probability that two base 
$A$ strings of length $m$ are mapped onto the same codeword is equal to $\left(\frac{1}{A}\right)^m$. 
Thus, the probability of a collision with $y$ is
\beq
p_{\rm col}=1-\left(1-\left(\frac{1}{A}\right)^m\right)^{k-1}.
\eeq
For small $2^{-l}$ we approximate
\beq
p_{\rm col}=1-\left(1-\frac{2^{-l}}{k}\right)^{k-1} 
\sim \ 2^{-l}(1-1/k)+O(2^{-2l})\ .
\label{arbsmall}
\eeq
This probability can be made arbitrarily small for only a relatively small
$l$.  Thus $O(\log_A k)$ random $g_i$s are sufficient to retrieve the 
information with arbitrarily low probability of error. 
It is clear that for negative $l$ the length of the codewords is not
sufficiently large to avoid collisions. A codeword length of $O(\log_A k)$ is 
thus necessary as well as sufficient. If the contents of the database 
are to be determined with certainty, the codeword length must 
be made larger.  A code with no collisions and with codeword
length $O(2 \log_A k)$ always exists (cf. the discussion of the birthday
problem \cite{feller}).

Our quantum algorithm to determine the contents of the database in
a single query with high probability makes use of this classical random 
coding construction. The random strings $g_i$, $i=1\ldots m$ are the 
generators of a group $C_A$. The multiplication rule for this
group is a digit-wise addition modulo $A$ and the identity element is the string $0$.
Members of $C_A$ can be written as
\beq
c(s) \in C_A \Rightarrow c(s)_k=\sum_i (g_i)_k s_i\; {\rm  mod}\; A
\eeq 
with $c(s)_k$ the $k^{\rm th}$ digit of a group element $c(s)$ and $s$ is a
base $A$ string of length $m$. Due to the linear independence of the 
generators, $C_A$ is a subgroup of $(\dubbelZ_A)^n$ with $A^m$ 
elements.  

For $c(s) \in C_A$ there is a one-to-one map between $c(s)$ and its encoding $z$ defined in 
(\ref{hash}).  This is true as
\beq
\forall_i \; \; c \in C_A,\; c \cdot g_i=0 \Leftrightarrow c=0 
\eeq 
which follows from the linear independence of the 
generators.

In the quantum algorithm we construct a state
\beq
\ket{\psi}=\frac{1}{\sqrt{m}}\sum_{s|c(s) \in C_A} \ket{s,c(s)} \otimes \frac{1}{\sqrt{A}}
\sum_{b=0}^{A-1}\omega_A^b \ket{b},
\eeq
with $\omega_A=e^{\frac{i 2\pi}{A}}$. The query results in the state
\beq
\ket{\psi_y}=\frac{1}{\sqrt{m}}\sum_{s|c(s) \in C_A} \omega_A^{-a(c(s),y)} \ket{s,c(s)} \otimes 
\frac{1}{\sqrt{A}}\sum_{b=0}^{A-1}\omega_A^b \ket{b}.
\eeq
We can write, using the encoding $z$ of $y$ defined in (\ref{hash}) 
\beq
a(c(s),y)=s \cdot  z.
\eeq
Thus we have
\beq
\inner{\psi_y}{\psi_{y'}}=\frac{1}{m}\sum_s \omega_A^{s \cdot  (z-z')} 
=\delta_{zz'}.
\eeq
If two different strings $y$ and $y'$ are mapped onto a different  
codeword, they are thus distinguishable by a measurement. The probability that 
this occurs (\ref{arbsmall}) can be made arbitrary small just as in the 
classical case since the encoding is the same.
In order to measure, we reverse the preparation steps and then we 
perform a Fourier transform over $(\dubbelZ_A)^n$
\beq
H_A\colon\;\ket{s} \rightarrow \frac{1}{\sqrt{m}} \sum_{z}\omega_A^{s\cdot z} \ket{z}.
\eeq
A measurement in the query basis determines $z$ and, with high probability, $y$.

The circuit used to implement the random coding is similar to that in
Figure \ref{fig2} but the XORS are replaced by summation base $A$ operators,
\beq
{\rm XOR}_A(a,b)=(a+b)\ {\rm mod}\  A\ 
\eeq
and their locations are according to the random queries.

The total quantum  running time is $2 + m n + T(n)$.  Here the basic unit of time
is an operation on an $A$-dimensional Hilbert space.  The classical time
using the same random codewords is $m/2 + m T(n)$.  Since $m$ is less than
$n$, this algorithm is better than the direct coin-weighing algorithm 
provided we are willing to tolerate a small chance of error bounded by $p_{\rm col}$.

\section{Discussion}

We have discussed the complexity of our quantum algorithms compared to 
a classical setup and shown that the quantum algorithms are faster 
in situations in which $T(n) > O(n)$.  
In problems where querying the database would occur repeatedly, a
bigger (real) separation between the classical computation time and the
quantum computation time could be achieved (see \cite{bv2} for an 
instance of such a problem). 

It is noteworthy that in the binary search problem 
in the classical case only the generators
of the Walsh functions are required, while the quantum algorithm
needs all the Walsh functions to achieve this speedup.  It would be
interesting to find out whether any speedup is possible if 
the database only responds to queries which are the generators.

We have chosen the quantum database $R_y$ as defined in
(\ref{requation}) to make a fair comparison with the classical
setting. A unitary $U_y$ could easily become more powerful as was
pointed out in \cite{machta}.  At its most general, a quantum database
could be defined by an arbitrary unitary transformation acting on an
input register and a hidden quantum state (the database). This has no
good classical analogue and might be be worthwhile to explore.

We would like to thank Charles H. Bennett, David P. DiVincenzo and Markus
Grassl for helpful discussions, and the Army Research Office and 
the Institute for Scientific Interchange, Italy, for financial support.  
B.M.T. would like
to thank Bernard Nienhuis and Paul Vitanyi for advice and encouragement.


\begin{thebibliography}{99}
\bibitem{nonuni} One could have a nonunitary quantum database
where the $X$ register is always reset to $\ket{0}$ to ensure
it gives no information on $Y$.  This would not matter to a classical
query where all the information is in $B$, but severely
cripples the quantum algorithm.  It is not known if such a database
will allow any improvement over classical algorithms.
\end{thebibliography}
\end{document}